\documentclass[english,prl,aps,twocolumn,10pt]{revtex4-2}

\usepackage{amsmath}
\usepackage{amssymb}
\usepackage{graphicx}
\usepackage[english]{babel}
\usepackage{array}
\usepackage{verbatim}
\usepackage[colorlinks=true, pdfstartview=FitV, linkcolor=blue, citecolor=blue, urlcolor=blue]{hyperref} % enable links

\def\C        {{$^{13}$C \/}}

\newcommand{\mr}[1]{\mathrm{#1}}
\newcommand{\unit}[1]{\,\mathrm{#1}}
\newcommand{\um}{\,\mu{\rm m}}
\newcommand{\us}{\,\mu{\rm s}}
\newcommand{\uT}{\,\mu{\rm T}}

\newcommand{\uW}{\,\mu{\rm W}}

\newcommand{\rtHz}{\sqrt{\mr{Hz}}}

\newcommand{\DB}{\Delta B}
\newcommand{\Bmax}{\Delta B_\mr{max}}
\newcommand{\tpi}{t_\pi}
\newcommand{\tlaser}{t_\mr{laser}}
\newcommand{\tset}{t_\mr{set}}
\newcommand{\Tc}{T_\mr{c}}
\newcommand{\Bc}{\ensuremath{B_{\mathrm{c2}}}}
\newcommand{\Tsample}{T_\mr{sample}}
\newcommand{\Truox}{\ensuremath{T_{\mathrm{RuO}_2}}}
\newcommand{\Truoxc}{\ensuremath{T_{\mathrm{RuO}_2}^\mr{(c)}}}
\newcommand{\Plaser}{P_\mr{laser}}
\newcommand{\Pmw}{P_\mr{mw}}
\newcommand{\dUdI}{\mr{d}U/\mr{d}I}

\begin{document}

%\title{Millikelvin scanning nitrogen-vacancy magnetometry}
\title{Scanning nitrogen-vacancy magnetometry down to 350\,mK}
%\title{A 350\,mK scanning NV magnetometer}
\author{P.~J.~Scheidegger$^{1,\dagger}$, S.~Diesch$^{1,\dagger}$, M.~L.~Palm$^1$, and C.~L.~Degen$^{1,2}$}
\email{degenc@ethz.ch}
\thanks{$^\dagger$These authors contributed equally.}
\affiliation{$^1$Department of Physics, ETH Zurich, Otto Stern Weg 1, 8093 Zurich, Switzerland.}
\affiliation{$^2$Quantum Center, ETH Zurich, 8093 Zurich, Switzerland.}

\begin{abstract}
We report on the implementation of a scanning nitrogen-vacancy (NV) magnetometer in a dry dilution refrigerator.  Using pulsed optically detected magnetic resonance combined with efficient microwave delivery through a co-planar waveguide, we reach a base temperature of 350\,mK, limited by experimental heat load and thermalization of the probe.  We demonstrate scanning NV magnetometry by imaging superconducting vortices in a 50-nm-thin aluminum microstructure.  The sensitivity of our measurements is approximately 3\,$\mu$T per square root Hz.  Our work demonstrates the feasibility for performing non-invasive magnetic field imaging with scanning NV centers at sub-Kelvin temperatures.
\end{abstract}

\date{\today}

\maketitle

%\textit{Introduction -- }
%
Imaging magnetic fields at the micro- to nanoscale is a powerful method for materials characterization and provides insights into a plethora of physical phenomena, ranging from magnetic ordering to the flow of electric currents~\cite{marchiori22}.  Over the past decade, nitrogen-vacancy (NV) centers in diamond scanning probes \cite{degen08apl,maletinsky12} have been introduced as versatile magnetic field sensors that are applicable over a broad temperature range.
At room temperature, researchers have used NV probes to map the magnetic structure in antiferromagnets~\cite{gross17,appel19,wornle21} and van-der-Waals (vdW) magnets \cite{fabre21}, and to image electronic transport in one-dimensional~\cite{chang17} and two-dimensional conductors~\cite{tetienne17,ku20,palm22}.
At cryogenic temperatures, initial experiments focused on the imaging of vortices in iron pnictide and cuprate high-temperature superconductors with critical temperatures of $\Tc = 30\unit{K}$ and $90\unit{K}$, respectively~\cite{pelliccione16,thiel16}.  More recently, vdW magnets have been studied at temperatures down to $4\unit{K}$ \cite{thiel19,sun21,song21}.  Cryogenic scanning NV magnetometers have also been used for spatial imaging of hydrodynamic electron flow at temperatures between $5-100\unit{K}$ \cite{jenkins20,vool21}.

There is considerable incentive to bring scanning NV magnetometry to sub-Kelvin temperatures.  Not only would this allow addressing interesting physics like superconductivity or magnetism in twisted bilayer graphene \cite{cao18,sharpe19} and in oxide interfaces \cite{bert11}, but it would also ideally complement existing low-temperature scanning probe techniques such as scanning superconducting quantum interference device (SQUID)  \cite{persky22}, scanning gate \cite{pelliccione_design_2013}, scanning tunneling and magnetic exchange force microscopy \cite{wiesendanger_spin_2009}.  The biggest challenges towards millikelvin NV magnetometry are microwave- and laser-induced heating, vibrational stability of the atomic force microscope (AFM), and the experimental complexity of cryogenic instrumentation.  Moreover, the low-temperature charge stability of shallow NV centers has been put into question \cite{rohner20,wise21} and the photo-luminescence (PL) emission \cite{rogers09,batalov09} and PL contrast \cite{wise21,ernst22} of NV centers have been found to decrease towards cryogenic temperatures.  It is therefore an intriguing technical and scientific question of whether scanning NV magnetometry can be robustly applied in the sub-Kelvin regime.

%%% In this work, ...
In this work, we demonstrate scanning NV magnetometry at temperatures down to $350\unit{mK}$.  Experiments are carried out at the bottom of a dry dilution refrigerator and employ pulsed optically detected magnetic resonance (ODMR) detection of the NV center to mitigate microwave- and laser-induced heating.  We find that NV centers are stable within the temperature range of our experiment ($0.35-3\unit{K}$) with no noticeable change in their spin and optical properties.  We demonstrate scanning magnetometry by imaging superconducting vortices in thin aluminum micro-discs exposed to weak ($0.4-1\unit{mT}$) magnetic bias fields.

\begin{figure*}[htp]
    \includegraphics{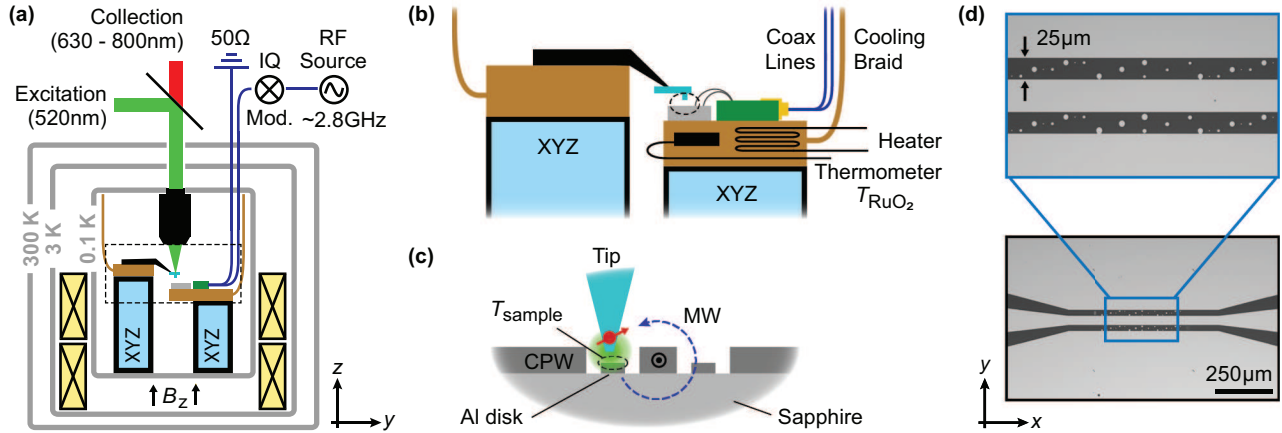}
    \caption{
		    (a) Schematic of the experimental setup. The scanning NV magnetometer is located at the bottom of a dry dilution refrigerator.  Optical excitation and detection use the clear central bore.  Microwaves are guided to the sample and dumped in an external $50\unit{\Omega}$ load using two semi-rigid coaxial lines. A vector magnet (yellow boxes) is used to apply the out-of-plane field $B_\mathrm{z}$.
			(b) Detail of the scanning magnetometer, indicating the relative positions of scanning tip and sample. The sample holder contains both, a resistive heater and a thermometer (\Truox). 
			(c) Close-up schematic of the scanning tip over the aluminum film. The NV center is shown by a red arrow. Microwave pulses (MW, blue arrow) are delivered through the co-planar waveguide (CPW). Refracted laser light is indicated in green. Due to heating effects, the local temperature $\Tsample$ can differ from thermometer readings (\Truox).
			(d) Optical micrograph of the patterned aluminum film.  The large structure is the CPW used for microwave delivery. The zoom-in shows the small ($1-5\unit{\um}$ diameter) discs placed in the gaps of the CPW that serve as test structures for scanning experiments.
		}
	\label{fig1}
\end{figure*}

%%% Experimental
%\textit{Experimental -- }
A schematic of our experimental setup is shown in Fig.~\ref{fig1}(a-c). The setup consists of a combined atomic force/confocal microscope (AFM/CFM, Attocube) rigidly suspended from the cold insert of a CF-CS110 dry dilution refrigerator (Leiden Cryogenics). The system features free-space access through a $8\unit{mm}$ central bore for the optical excitation of the NV spin and collection of the PL signal. The objective is fixed in space while both the sample and the diamond tip are mounted on stacks of nano-positioners and scanners for movement along all three spatial axes.  The microwave (MW) drive signal needed to manipulate the NV center spin is guided to the sample through a pair of coaxial lines (a combination of semi-rigid \textsc{Coax Co.} SC‐219/50‐SB‐B, SC‐119/50 CN‐CN and superconducting SC‐219/50 NbTi‐NbTi, % off-the-shelf semi-flexible %\textsc{Fairview Microwave} FM-F086; 
off-the-shelf semi-flexible coax cables and flexible extensions to connect to the sample holder). % that are thermalized at each cooling stage of the cryostat.  
A superconducting vector magnet (American Magnetics Inc.) is used to apply a small out-of-plane bias field $B_\mathrm{z}$. As a dry pulse-tube system, our refrigerator introduces noticeable vibrations between scanning tip and sample. To retain a high beam-pointing stability for our confocal microscope, we built a rigid system rather than employing commonly used spring-suspension for vibrational damping \cite{dewit19}.  We did not calibrate the vibrations in our cryostat but estimate them to be of order of tens of nanometers, similar to comparable rigid dry systems \cite{low21}.

We glue our sample onto a copper sample holder, which also contains a resistive heater and a calibrated ruthenium oxide (RuO$_2$) thermometer ($\Truox$), as is shown in Fig.~\ref{fig1}(b).  A cooling braid serves as a direct thermal link between the sample holder and the cold finger of the insert.  The insert is in turn thermally anchored to the mixing chamber by a spring-loaded mechanical contact arm (Leiden Cryogenics).  We measure a base temperature of $\Truox = 140\unit{mK}$ for the sample holder without applying experimental heat load.  The base temperature of the mixing chamber is significantly lower ($\sim 25\unit{mK}$); we attribute the difference to limited thermal anchoring of the cold insert combined with heat transmitted through the wiring and radiation along the bore.

Our experiment makes use of two nanofabricated structures, including the sample and the scanning tip.  The sample consists of an aluminum film structured into a co-planar waveguide (CPW) for microwave delivery as well as several micron-sized discs for demonstrating scanning magnetometry.  Aluminum is an ideal material for our demonstration because it becomes superconducting below $\Tc\sim 1.25\unit{K}$ \cite{meservey71}, allowing us to use the Meissner effect as a magnetic hallmark.  Both the CPW and discs are lithographically patterned on a sapphire substrate in a two-step process via e-beam evaporation at room temperature at a rate of $0.4\unit{nm\,s}^{-1}$.  We first define the impedance matched, tapered CPW with a thickness of $150\unit{nm}$.  We then add discs of $50\unit{nm}$ nominal thickness and various diameters ($1-5\unit{\um}$) inside the CPW gaps.  See Fig.~\ref{fig1}(c) for a schematic side-view of the CPW and the discs. An optical micrograph of the sample is shown in Fig.~\ref{fig1}(d).
The scanning tip is made from a monolithic block of single-crystal diamond (\{100\} surface orientation) using a series of dry etching steps~\cite{maletinsky12,chang16thesis,welter22}.  The diamond tip is attached via a silicon handle structure to a quartz tuning fork oscillator for distance control.  NV centers are created by low energy ion implantation ($7\unit{keV}$).  We use tips from two batches in the course of this study; all tips are from QZabre AG~\cite{qzabre}.

%%% Pulsed ODMR
%
\begin{figure}
    \includegraphics{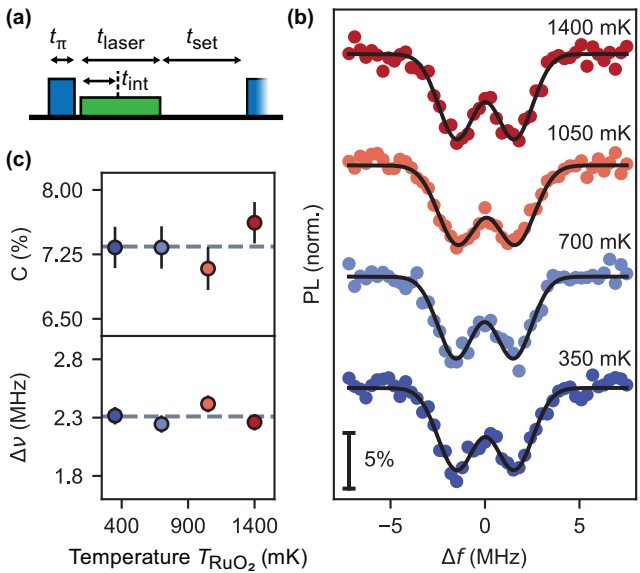}
    \caption{
		    (a) Pulse sequence used for ODMR detection.  Microwave $\pi$-pulses are drawn in blue, green represents the laser excitation, and the sketched heights indicate the respective powers. The PL is detected continuously and integrated in post-processing over a window of duration $t_\mathrm{int}$. We adjust pulse durations $\tpi$ and $\tlaser$ and settling time $\tset$ to both maximize sensitivity and minimize thermal load.
			(b) Pulsed ODMR spectra for various stage temperatures $\Truox = 0.35-1.4\unit{K}$.  Horizontal axis is frequency of microwave excitation and vertical axis is normalized PL (spectra are offset for clarity).  Two dips are visible for all spectra corresponding to the two hyperfine lines of the $^{15}$N nuclear spin.  A double-Gaussian is fitted to the data (black line). Here, $\tpi = 500\unit{ns}$, $\tlaser = 1.5\unit{\us}$ and $\tset = 1.5\unit{\us}$.
			(c) Plots of the fitted contrast $C$ (top panel) and line width $\Delta \nu$ (bottom panel) against temperature. Error bars represent fit uncertainties, the dashed grey lines represent the arithmetic mean.
		}
	\label{fig2}
\end{figure}
%

\begin{figure*}[htp]
    \includegraphics{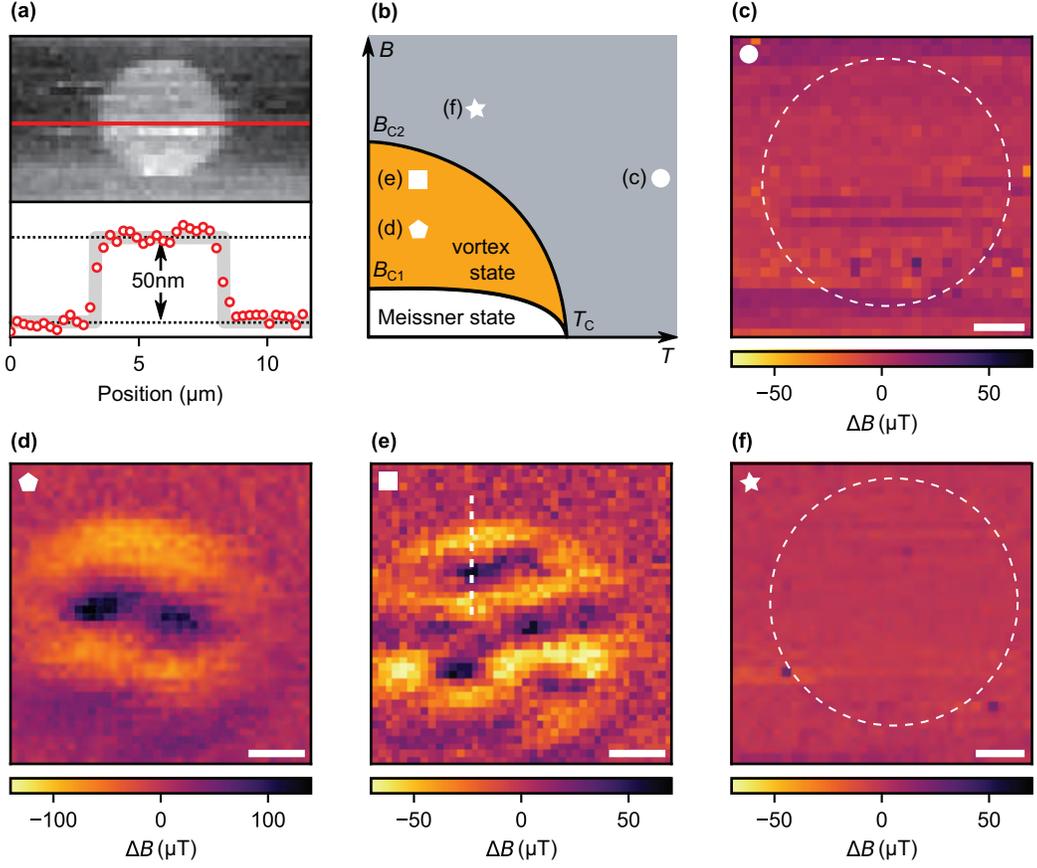}
    \caption{
		    (a) AFM topography of a 50-nm-thick, 5-$\um$-diameter aluminum disc and a linecut along the red line. The vertical root-mean-square stability of the linecut is $\sim 2.5\unit{nm}$ despite the presence of pulse tube vibrations.
			(b) Sketch of the phase diagram of a type II superconductor. White symbols indicate temperature and external magnetic field of the following four magnetometry scans:
			(c) Magnetic field map taken at $\Truox=3\unit{K} > \Tc$ and $B_\mathrm{z}=1\unit{mT}$; with pixel size $a_\mathrm{px}=208\unit{nm}$ and dwell time $t_\mathrm{px}=30\unit{s}$.
			(d) Magnetic field map taken at $\Truox=0.35\unit{K}$ and $B_\mathrm{z}=0.4\unit{mT}$; $a_\mathrm{px}=133\unit{nm}$ and $t_\mathrm{px}=30\unit{s}$.
			(e) Magnetic field map taken at $\Truox=0.35\unit{K}$ and $B_\mathrm{z}=1.0\unit{mT}$; $a_\mathrm{px}=133\unit{nm}$ and $t_\mathrm{px}=30\unit{s}$.
			(f) Magnetic field map taken at $\Truox=0.69\unit{K}$ and $B_\mathrm{z}=6\unit{mT} > B_{\mathrm{C2}}$; $a_\mathrm{px}=185\unit{nm}$ and $t_\mathrm{px}=60\unit{s}$. 
			The dashed circles in (c) and (f) indicate the disc position determined from the AFM height signal. No magnetic features are visible, because superconductivity has been suppressed.
			Dark features in (d) and (e) reflect superconducting vortices where the field is enhanced.
			The dashed line in (e) indicates the position of the line scans presented in Fig.~\ref{fig4}.
			Data are taken on multiple, nominally identical discs; disc (d-e) has a slight perforation at the bottom left corner.
			The NV-to-sample distance is ca. $110\unit{nm}$.
			Scale bars are $1 \unit{\um}$.
				}
	\label{fig3}
\end{figure*}

\begin{figure*}[htp]
    \includegraphics{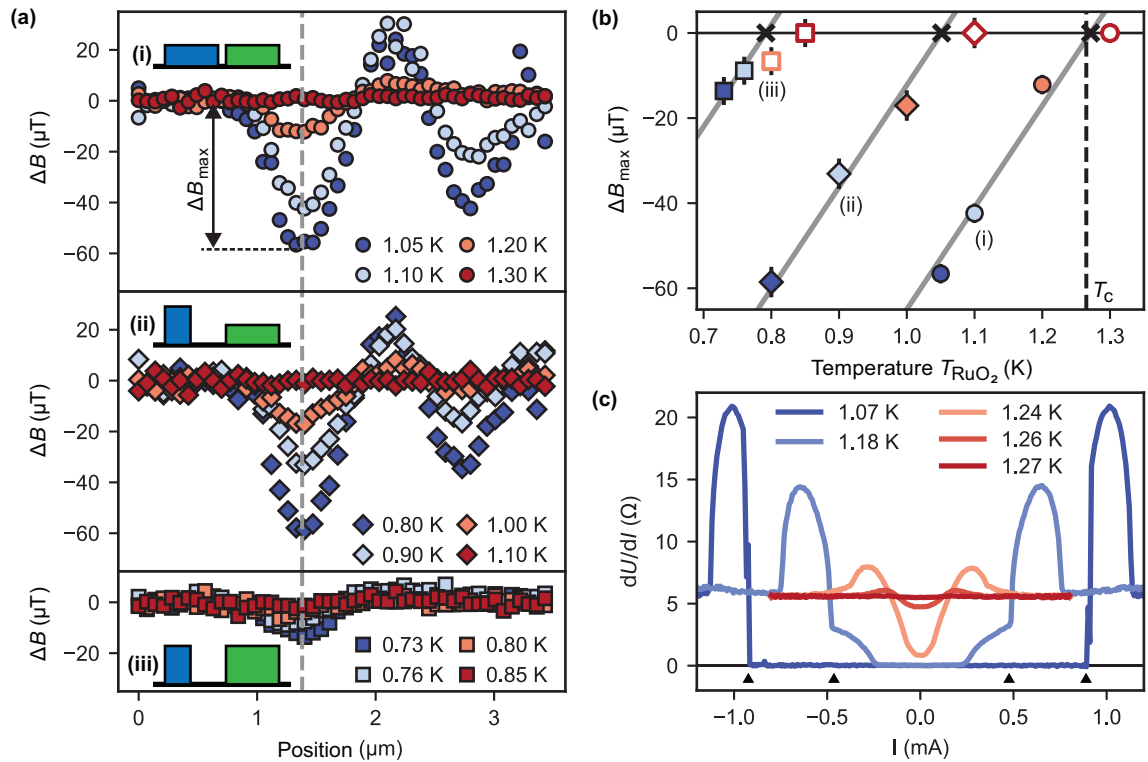}
    \caption{
		    (a) Magnetic line scans taken along the dashed line in Fig.~\ref{fig3}(e). Colors indicate stage temperature $\Truox$. Datasets (i)-(iii) are recorded using different settings with increasing microwave and laser power, as indicated by the sequence drawings (see main text for values).  All data are recorded in an $1\unit{mT}$ out-of-plane bias field.
			(b) Peak signal $\Bmax$ plotted as a function of $\Truox$ for all datasets, with error bars representing the pixel-to-pixel variations of the signal-less, high temperature scans.  Solid lines are linear fits (hollow data points are excluded from the fit).  A common slope parameter is used for all fits.  The zero crossings (black crosses) reflect the critical stage temperatures $\Truoxc$ at which the superconductivity vanishes, indicating that the sample temperature $\Tsample=\Tc$ at this stage temperature. The $\Truoxc$ determined from the fit are $\approx1.27\unit{K}$ (i), $1.05\unit{K}$ (ii) and $0.79\unit{K}$ (iii)
			, where (i) is in excellent agreement with transport measurements (dashed line, details in (c)).
			(c) Electric transport measurements recorded on a second microstrip similar to the one shown in Fig.~\ref{fig1}(d).  Shown is a three-point measurement of the differential resistance $\dUdI$ plotted as a function of the applied current $I$, for different stage temperatures $\Truox=1.07-1.27\unit{K}$.  The sharp drops in $\dUdI$ (triangles) indicate the critical current. A finite contact resistance ($\sim 18.4\unit{\Omega}$) is subtracted from all curves.
			}
	\label{fig4}
\end{figure*}

%\textit{Results -- }
We start experiments by recording a set of ODMR spectra as a function of stage temperature $\Truox = 0.35-1.4\unit{K}$.  We use a pulsed ODMR scheme (Fig.~\ref{fig2}(a) and Ref.~\onlinecite{dreau11}) to both decrease the line width of the spin resonance and reduce the duty cycle of laser and microwave irradiation.
Fig.~\ref{fig2}(b) shows representative spectra from tip \#2. To extract the transition frequency, PL contrast and line width, we fit a double-Gaussian to the ODMR spectra, which accounts for the hyperfine-splitting caused by the the $^{15}$N nuclear spin. We choose a Gaussian over a Lorentzian line shape because our spectral width is dominated by the \C bath rather than the microwave excitation \cite{dreau11}. Our measurements show that the PL contrast $C$ and resonance line width $\Delta \nu$ are independent of temperature between $0.35-1.4\unit{K}$. The fitted PL contrast is constant to within $0.5\%$ across all temperatures, and the line width varies by less than $200\unit{kHz}$, as plotted in Fig.~\ref{fig2}(c). The PL contrast of a single hyperfine-transition for tip \#2 ($\sim 7.3\%$) is comparably low; other tips showed contrast values of up to $15\%$. On average, we observe a $\approx 30\unit{\%}$ reduction compared to the values measured at room temperature \cite{ernst22}.  (Note that since we are only driving one hyperfine transition, PL contrast values are roughly one-half compared to those measured by standard ODMR \cite{dreau11}.)
Furthermore, none of the 10 scanning tips tested at temperatures of $3\unit{K}$ and below showed unexpected bleaching or loss of contrast.  Noteably, tip \#1, which was used for all scanning magnetometry experiments (Fig.~\ref{fig3} and \ref{fig4}), has been in continuous operation at cryogenic conditions for more than one month without showing signs of deterioration.  Overall, Fig.~\ref{fig2} demonstrates that NV centers in scanning tips can be employed down to at least $350\unit{mK}$ with no adverse effects.

We proceed to demonstrate scanning magnetometry on the aluminum discs above and below the superconducting transition temperature $\Tc$ at varying out-of-plane magnetic fields $B_\mathrm{z}$. To record a magnetic field map,  we scan the tip over the sample in contact mode and acquire a pulsed ODMR spectrum at each pixel. We obtain the local stray field $\DB$ from the frequency shift of the resonance (determined by a least-squares fit, see Fig.~\ref{fig2}(b)) compared to a far-away reference location, and converting the frequency shift to units of Tesla using the NV spin's gyromagnetic ratio ($\gamma = 2\pi\times 28\unit{GHz/T}$) \cite{welter22}.  Note that the frequency shift is proportional to magnetic field component along the NV symmetry axis, which lies at an $\sim 55^\circ$ angle with respect to the out-of-plane direction for our tips.

%%% Al disc scans
%

Over the course of our study, we have recorded magnetic images of multiple aluminum discs of the same size, which all show the same magnetic behaviour. In Fig.~\ref{fig3}, we present magnetometry images taken on two representative aluminum discs, together with the sample topography (Fig.~\ref{fig3}(a)) and a schematic phase diagram (Fig.~\ref{fig3}(b)).  When performing magnetometry at $3\unit{K}$, which lies above $\Tc\approx 1.25\unit{K}$, we record no change in the magnetic field above the disc (Fig.~\ref{fig3}(c)).  Upon cooling the sample to the fridge's base temperature ($\Truox=0.35\unit{K}$) a spatially diverse magnetic pattern emerges, indicating the onset of superconductivity in the sample (Fig.~\ref{fig3}(d)).
Notably, we observe the formation of superconducting vortices on the aluminum disc: while bulk aluminum is a type I superconductor, which would display complete expulsion of the magnetic field (Meissner effect), thin films of aluminum have been reported to exhibit type II superconducting properties including the formation of a vortex state \cite{tinkham96}.  At $B_\mathrm{z}=0.4\unit{mT}$, two vortices are visible (Fig.~\ref{fig3}(d)); once we raise the field to $1.0\unit{mT}$, the number of vortices increases to approximately 8 (Fig.~\ref{fig3}(e)).  At the same time, the vortex cores shrink to approximately half their initial size, and the peak magnitude of the magnetic signal decreases from $140\unit{\uT}$ to $70\unit{\uT}$.  Both effects -- the reduction of the core size \cite{fente16} and the continuous decrease in magnetic susceptibility with temperature and field \cite{cochran58} -- correspond to the expected behavior for type II superconductors. In both scans, the vortices appear to have a spatial asymmetry. This is partially explained by the projection of the magnetic field onto the tilted NV axis \cite{thiel16}. However, we observe some additional elongation along the x-direction for which we have no ready explanation and which might be caused by vibrations inside our cryostat. We further observe a clustering of several of the vortices in Fig.~\ref{fig3}(e), possibly due to pinning at defect sites such as grain boundaries \cite{kirtley10}. The apparent merging of vortices is likely explained by the asymmetry combined with the close spacing of the vortices.
As we increase $B_\mathrm{z}$ to $6\unit{mT}$, which is well above the upper critical magnetic field $\Bc$ of our sample, superconductivity is completely suppressed and we again record a uniform magnetic signal above the disc (Fig.~\ref{fig3}(f)). 

In this final scan, the settings were chosen to yield particularly narrow hyper-fine lines, which allows us to estimate the achievable magnetic-field sensitivity.  Normalizing the pixel-by-pixel variation to a one-second integration time, we obtain a sensitivity of about $14\unit{\uT/\rtHz}$. This number is likely affected by the already sizeable off-axis field. Evaluating the line scans discussed further below (Fig.~\ref{fig4}(a)), which were taken with comparable frequency resolution but at a smaller $1\unit{mT}$ field, yields a considerably improved sensitivity of $3\unit{\uT/\rtHz}$.

%%% microwave and laser heating

We finally investigate the influence of microwave and laser pulses on the sample temperature.  Since the RuO$_2$ thermometer is located a few mm away from the sample (see Fig.~\ref{fig1}(c)), the true sample temperature $\Tsample$ is likely higher than the measured temperature $\Truox$.  To characterize the local heating, we perform NV magnetometry line scans across the edge of the aluminum disc (dashed line in Fig.~\ref{fig3}(e)) and observe the peak field $\Bmax$ near the sample edge.  Because $\Bmax$ is approximately proportional to temperature near $\Tc$~\cite{cochran58}, we can use the peak field as an indicator for the sample temperature $\Tsample$.

Fig.~\ref{fig4}(a) presents line scans as a function of stage temperature $\Truox$.  We regulate the stage temperature via the built-in sample heater (Fig.~\ref{fig1}(c)) and a PID controller.  We investigate three laser and microwave power settings, from low to high:
(i) $\Plaser = 50\unit{\uW}$ and $\Pmw = 70\unit{\uW}$,
(ii) $\Plaser = 50\unit{\uW}$ and $\Pmw = 300\unit{\uW}$, and
(iii) $\Plaser = 140\unit{\uW}$ and $\Pmw = 300\unit{\uW}$.
The given powers are time-averaged over the respective measurement sequence.  Laser powers are measured at room temperature in front of the objective.  Microwave powers reflect the nominal input power minus $5\unit{dB}$ to account for the attenuation of the cold coax lines. 

In Fig.~\ref{fig4}(b), we plot the peak field $\Bmax$ as a function of $\Truox$ for all datasets (i)-(iii).  By applying linear fits (solid lines), we extrapolate the critical stage temperature $\Truoxc$ above which superconductivity vanishes.  We use a common slope parameter when fitting the three datasets, and we exclude points that do not carry signal, \textit{i.e.}, points that lie above $\Truoxc$.  The so-determined critical stage temperature $\Truoxc \leq \Tc$ is in general lower than the critical temperature of the superconductor $\Tc$ due to the temperature difference between RuO$_2$ thermometer and sample.  For the lowest power setting (i), however, we observe that the extrapolated $\Truoxc \approx 1.27\unit{K}$ is very close to the $\Tc = 1.25\unit{K}$ expected for 50-nm-thick aluminum films, for which $\Tc$ approaches bulk values \cite{meservey71}.  Therefore, for (i), the local heating by microwave and laser pulses is negligible.

To corroborate these results and estimate the actual $\Tc$ of our sample, we perform complementary electrical transport measurements on a separate aluminum film of similar thickness ($\sim 60\unit{nm}$).  Fig.~\ref{fig4}(c) shows the differential electrical resistance $\dUdI$ as a function of applied current $I$, at $\Truox=1.07-1.27\unit{K}$.  The sharp jumps in $\dUdI$ (triangles) indicate the critical current $I_\mr{c}$ below which superconductivity sets in.  With increasing temperature, the critical current is reduced and the $\dUdI$ curve becomes flatter, reflecting that the film's temperature is approaching $\Tc$.  Any signs of superconductivity vanish at $\Truox = 1.27\unit{K}$, in excellent agreement with the literature and magnetometry values.

%%% Conclusions
%\textit{Conclusions -- }
%
In summary, our work extends scanning NV magnetometry to sub-Kelvin temperatures.  Although our base temperature ($\sim 350\unit{mK}$) and sensitivity ($\sim 3\unit{\uT/\rtHz}$) are above those reached by state-of-the-art scanning SQUID microscopes~\cite{uri20nphys,low21}, there is considerable leeway for improving both figures of merit.  Base temperature can be lowered by improved probe thermalization, more efficient microwave delivery~\cite{fuchs09}, and possibly using low power, resonant optical readout~\cite{robledo11nature}.  The sensitivity can be improved using ac magnetometry~\cite{vool21,palm22} and gradiometry~\cite{huxter22} detection where demonstrated noise levels are below $100\unit{nT/\rtHz}$.  A remaining technical issue is spurious mechanical vibrations from the cooling circuit of the dry dilution refrigerator.  With these challenges addressed, scanning NV magnetometry shows exciting promise for the non-invasive imaging of nanoscale transport, persistent ring, edge and vortex currents~\cite{bluhm09current, uri20nphys}, superconductivity~\cite{cao18, bert11} and magnetic structures with phase transition temperatures of a few hundred millikelvin and above.

%%%%%%%%%%%%%% Acknowledgments

\vspace{0.5cm}\textbf{Acknowledgments -- }
The authors thank S. Ernst, K. Herb, W. Huxter, and P. Welter for fruitful discussions, and U. Grob and the staff of the FIRST lab clean-room facility for technical support.
This work was supported by the European Research Council through ERC CoG 817720 (IMAGINE), the Swiss National Science Foundation (SNSF) through Project Grant No. 200020\_175600 and through the NCCR QSIT, a National Centre of Competence in Research in Quantum Science and Technology, Grant No. 51NF40-185902, and the Advancing Science and TEchnology thRough dIamond Quantum Sensing (ASTERIQS) program, Grant No. 820394, of the European Commission.

%\textit{Author contributions --} C.L.D, P.J.S. and S.D. conceived the experiment. P.J.S. and S.D. carried out all experiments and performed the data analysis. M.L.P. fabricated the sample. P.J.S., D.S. and C.L.D. wrote the manuscript. All authors discussed the results.

%%%%%%%%%%%%% Statements

\vspace{0.5cm}\textbf{Conflict of interest -- }
The authors have no conflicts to disclose.

\vspace{0.5cm}\textbf{Data availability statement -- }
The data that support the findings of this study are available from the corresponding author upon reasonable request.

\bibliography{library}

\end{document}